\documentclass[notitlepage,aps,pra,twocolumn,longbibliography,superscriptaddress]
{revtex4-1}
\usepackage{graphicx}
\usepackage{amsmath}
\usepackage{amssymb}
\usepackage{comment}
\usepackage[colorlinks, allcolors=blue]{hyperref}
\usepackage[all]{hypcap}
\usepackage[mathlines]{lineno}
\usepackage{physics}
\usepackage{wrapfig}
\usepackage{svg}
\usepackage{lipsum}
\usepackage{ulem}
\usepackage{siunitx}
\DeclareUnicodeCharacter{2212}{-}

\newcommand{\pref}[2]{\hyperref[#1]{\ref{#1}(#2)}}
\newcommand{\preff}[2]{\hyperref[#1]{\ref{#1}#2}}
\newcommand{\eqpref}[1]{\hyperref[#1]{(\ref{#1})}}

\newcommand{\squig}{{\raise.17ex\hbox{$\scriptstyle\sim$}}}

\begin{document}
\title{Two dimensional momentum state lattices}
\author{Shraddha Agrawal}
\affiliation{Department of Physics, University of Illinois at Urbana-Champaign, Urbana, IL
61801-3080, USA}
\author{Sai Naga Manoj Paladugu}
\affiliation{Department of Physics, University of Illinois at Urbana-Champaign, Urbana, IL
61801-3080, USA}
\author{Bryce Gadway}
\email{bgadway@illinois.edu}
\affiliation{Department of Physics, University of Illinois at Urbana-Champaign, Urbana, IL
61801-3080, USA}
\date{\today}

\begin{abstract}
Building on the development of momentum state lattices (MSLs) over the past decade, we introduce a simple extension of this technique to higher dimensions. Based on the selective addressing of unique Bragg resonances in matter-wave systems, MSLs have enabled the realization of tight-binding models with tunable disorder, gauge fields, non-Hermiticity, and other features. Here, we examine and outline an experimental approach to building scalable and tunable tight-binding models in two dimensions describing the laser-driven dynamics of atoms in momentum space. Using numerical simulations, we highlight some of the simplest models and types of phenomena this system is well-suited to address, including flat-band models with kinetic frustration and flux lattices supporting topological boundary states. Finally, we discuss many of the direct extensions to this model, including the introduction of disorder and non-Hermiticity, which will enable the exploration of new transport and localization phenomena in higher dimensions.
\end{abstract}
\maketitle

\section{Introduction}


Atomic, molecular, and optical (AMO) systems are naturally free from intrinsic disorder and admit a high level of control and tunability. As such, a wide variety of AMO platforms have been utilized for the engineering of synthetic quantum matter, utilized in particular for the exploration of novel phenomena related to, \textit{e.g.}, condensed matter physics. The natural synergy between AMO systems and explorations of condensed matter phenomena range from the direct emulation of Hubbard model physics based on atoms in optical lattices~\cite{JAKSCH200552,esslinger-hubbard} to explorations of topological band phenomena in a variety of photonics experiments~\cite{Top-Phot-Ozawa}.
%

There have been continued efforts to expand the atomic and photonic toolboxes to enable the engineering of different kinds of lattice models for quantum simulation experiments. In this vein, techniques based on synthetic dimensions~\cite{ozawa2019topological} - where transport is explored not in real space but rather in sets of internal states or other auxiliary degrees of freedom - have garnered attention as of late for their ability to explore the physics of electronic matter in large electromagnetic fields. This topic area, most often associated with quantum Hall physics, gauge fields, and topology, has been hard to study through traditional AMO techniques, owing in part to the neutrality of atoms and photons. In synthetic dimensions, complex hopping and gauge fields are more natural to realize, because hopping terms often involve the absorption or emission of a photon. Platforms that feature multiple synthetic dimensions spanned by independent degrees of freedom are particularly exciting, as they offer the ability to engineer gauge fields fully in synthetic dimensions~\cite{Dutt-multiple}.

Here, we propose a general approach to creating two-dimensional synthetic lattices formed from laser-coupled atomic momentum states. This approach offers the ability to realize highly tunable tight-binding Hamiltonians that can be used for the exploration of a range of phenomena related to lattice transport. The local control of nearly all terms in the engineered models lends itself naturally to explorations of artificial gauge fields, disordered and quasiperiodic lattices, kinetic frustration, and in particular phenomena arising from the confluence of these ingredients as well as nonlinear atomic interactions.

This paper is organized as follows: In Sec.~\ref{1Dmsls}, we review the experimental approach to creating momentum state lattices (MSLs) in one dimension based on state-preserving Bragg transitions. In Sec.~\ref{2Dmsls}, we discuss the extension of this approach to higher dimensions, detailing a specific protocol for engineering MSLs in two dimensions. We discuss the lattice models that may be easily constructed based on this protocol using only first-order Bragg transitions, using simulations to motivate specific examples of exploring flat-band physics and topological spectra and edge states. In Sec.~\ref{extensions}, we discuss extensions based on parameter variation, higher-order Bragg processes, atomic interactions, and more. Finally, we conclude in Sec.~\ref{conc}.

\section{MSLs in 1D and quasi-1D}
\label{1Dmsls}

We begin by reviewing the atom-optics theory underlying the engineering of MSLs in one dimension (1D), as well as a discussion of extensions to quasi-1D systems (\textit{i.e.}, ladder-like geometries). This will provide a basis for the discussion of engineering MSLs in higher dimensions.

\subsection{MSLs in 1D}

We consider a generic system of two-level atoms of mass $M$, having a single internal ground (excited) state $\ket{g}$ $(\ket{e})$ with an energy $\hbar \omega_{g(e)}$ (and a ground-excited energy separation $\hbar\omega_{eg} = \hbar\omega_{e} - \hbar\omega_{g}$). These atoms and their interaction with a driving electric laser field $\textbf{E}$, neglecting spontaneous emission, are described in the dipole approximation by the single-particle Hamiltonian
\begin{equation}
\hat{H} = \frac{\hat{\textbf{p}}^2}{2M} + \hbar\omega_{g} \ket{g}\bra{g} + \hbar\omega_{e} \ket{e}\bra{e} - \textbf{d}\cdot\textbf{E} \ ,
\label{dip}
\end{equation}
where $\textbf{p}$ is the atomic free-particle momentum and $\textbf{d} = - |e|\textbf{r}$ is the atomic dipole operator, with $\textbf{r}$ a vector pointing from the nucleus to the valence electron coordinate. We now consider the electric field $\textbf{E}$, composed of two contributions - a right-traveling field $\textbf{E}^+(\textbf{r},t)$, given by 
\begin{equation}
     \textbf{E}^+(\textbf{r},t) = \textbf{E}^+ \cos(\textbf{k}^+ \cdot \textbf{r} − \omega^+ t + \phi^+)
\end{equation}
and a left-traveling field with a number of discrete frequency components, given by
\begin{equation}
     \textbf{E}^-(\textbf{r},t) =\sum_j \textbf{E}^-_j \cos(\textbf{k}_j ^- \cdot \textbf{r}-\omega_j^- t +\phi_j^-) \ .
\end{equation}
 
We assume that the fields propagate along the $\mathbf{e}_x$ axis and that the fields are nearly monochromatic, such that $\textbf{k}^+ = k\mathbf{e}_x$ and $\textbf{k}_j^- \simeq -k\mathbf{e}_x$ for all \textit{j}, where $k = 2\pi / \lambda$ is the wave vector of the lattice light of wavelength $\lambda$. 
For each frequency component of the electric field, the resonant Rabi couplings are $\Omega^+ = -\bra{e}\textbf{d}\cdot\textbf{E}^+ \ket{g}/ \hbar$ and $\Omega^-_j=-\bra{e}\textbf{d}\cdot\textbf{E}_j^- \ket{g}/ \hbar$. Without loss of generality, we may assume that $\textbf{E}^+ = \textrm{E}^+ \hat{z}$ and $\textbf{E}_j^- = \textrm{E}_j^- \hat{z}$, \textit{i.e.}, that the strength of the two fields does not vary over the region considered, and that they have a common polarization along the $z$ axis.
Finally, we restrict to the case that all of the applied frequencies are detuned from the $\ket{g}$ $\leftrightarrow$ $\ket{e}$ atomic resonance 
by a nearly common \textit{large} detuning $\Delta \equiv \omega_{eg}-\omega^+ \simeq \omega_{eg}-\omega_j^-$, such that essentially no laser-driven dynamics occur at first-order. 

Thus, with an experimental laser detuning $\Delta$ that is much greater than all relevant Doppler shifts and resonant Rabi couplings, the excited state $\ket{e}$ will acquire negligible population and can be effectively traced out. The light-atom interactions can then be described by the effective two-photon Bragg processes that change the atomic momenta by $\pm\hbar \textbf{k}_\textrm{eff}=\pm 2 \hbar k \mathbf{e}_x$ while preserving the ground internal state~\cite{Gould-Bragg,martin1988,kozuma1999}. These momentum transfers result from the virtual absorption of a photon from the right-traveling field and stimulated emission into the left-traveling field (and vice versa). Assuming all atoms begin with essentially zero momentum, momentum changes in discrete steps of $\pm \hbar \textbf{k}_\textrm{eff}$
suggest a description in terms of a basis of plane-wave momentum states labeled as $\ket{n}$ having momentum $\textbf{p}_n=2 n \hbar k \mathbf{e}_x$.

This discrete set of momentum modes forms an effective lattice in a so-called ``synthetic dimension''~\cite{ozawa2019topological}, where the 
momentum modes play the role of lattice sites and the Bragg transitions lead to a kind of laser-assisted tunneling between sites.
Ignoring spatial (trap) confinement and atomic interactions, the atoms can be assumed to have a purely quadratic dispersion, 
hence the mode-dependent kinetic energy $E_n=\textbf{p}_n ^2/2M=n^2 4 E_R$, where $E_R$ is the one-photon recoil energy given by $\hbar^2 k^2 /2M$. The full Hamiltonian in this plane-wave basis is
\begin{equation}
    H(t)=\sum_n E_n \ket{n}\bra{n}+\chi(t) \ket{n+1}\bra{n}+\chi^{*}(t)\ket{n}\bra{n+1} \ ,
\label{fullsim1D}
\end{equation}
where
\begin{equation}
    \chi(t)= \sum_j \hbar \tilde{\Omega}_j e ^{i \phi_j} e^{-i \Delta\omega_j t} \ .
\end{equation}
Here, $\chi(t)$ is the common, time-dependent off-diagonal coupling term that leads to transitions, or ``hopping,'' between the momentum orders. In terms of its spectral composition, $\chi$ can be seen to host a comb of drive frequencies $\Delta\omega_j = (\omega^+ - \omega^-_j)$, which is effectively derived from the spectral comb written onto the field $\textbf{E}^-$. The individual strengths and phases of these tones are given by $\tilde{\Omega}_j=\Omega^+ \Omega^-_j/ 2 \Delta$ and $\phi_j=\phi^+ - \phi^-_j$, respectively.

In the simplest scenario, each tone with index $j$ of the drive $\chi$ is associated with a unique first-order Bragg transition, \textit{e.g.}, $\ket{n_j} \leftrightarrow \ket{n_j+1}$.
This is enforced by having the frequency of a given tone, $\omega^+ - \omega^-_j$, set to be equal or nearly equal to a given two-photon Bragg resonance frequency $\tilde{\omega}_{n_j}$.
Here, $\hbar\tilde{\omega}_n$ is the energy difference between two neighboring momentum orders $n$ and $n+1$, relating to the Doppler frequency shift of the transition $\ket{n} \leftrightarrow\ket{n+1}$.
The free particle dispersion is quadratic, and hence its linear first derivative relates to the Doppler frequency shift
\begin{equation}    
\tilde{\omega}_n \equiv \frac{\textbf{p}_n\cdot\textbf{k}_\textrm{eff}}{M} + \frac{ \hbar\lvert\textbf{k}_\textrm{eff} \rvert^2}{2M}=(2n+1)4 E_R/\hbar
\label{res1Da}
\end{equation}
or alternatively, the energy difference between two neighbouring orders $\ket{n}$ and $\ket{n+1}$ is
\begin{equation}
\hbar\tilde{\omega}_n \equiv E_{n+1}-E_{n}=(2n+1)4 E_R \ .
\label{res1Db}
\end{equation}
Eqs.~\ref{res1Da} and \ref{res1Db} define the two-photon Bragg resonance condition for the transition $\ket{n} \leftrightarrow \ket{n+1}$.

Up to small two-photon detunings $\zeta_j$, each frequency drive tone is set to address one of the Bragg resonances, \textit{i.e.}, $\omega^+ - \omega^-_j = \tilde{\omega}_{n_j} - \zeta_j$.
Spectral engineering of an effective time-independent MSL Hamiltonian, $H_\textrm{eff}$, then follows by composing the frequency tones of $\chi$ such that they build up an MSL in a link-by-link fashion.
Unique spectral addressing is ensured by restricting all of the $\tilde{\Omega}_j$ and $\zeta_j$ terms of the driving field $\chi$ to be small compared to the frequency spacing between first-order Bragg resonances, $8 E_R/\hbar$. 
Following a transformation to the interaction picture and a rotating wave approximation, as well as a re-absorption of the $\zeta_j$ terms onto the diagonal by a redefinition of the creation and annihilation operators, the full Hamiltonian in Eq.~\ref{fullsim1D} can be simply recast \cite{gad2015,meier2016} by the first-order effective MSL Hamiltonian
\begin{equation}
    H_\textrm{eff}/\hbar=\sum_n \varepsilon_n \hat{c}^\dagger_{n}\hat{c}_n + \sum_n J_n(e^{i\varphi_n}\hat{c}^\dagger_{n+1}\hat{c}_n + \text{h.c.}) \ .
\end{equation}
To first order, the spectral properties of $\chi$ can be directly associated with the terms of $H_\textrm{eff}$ as $\zeta_j = \varepsilon_{n_{j+1}} - \varepsilon_{n_{j}}$, $\tilde{\Omega}_j = J_{n_j}$, and $\phi_j = \varphi_{n_j}$. To note, more refined descriptions can be made by also accounting for higher-order corrections \cite{gou2020}, especially important when incorporating higher-order Bragg transitions \cite{faan2018,gou2020}.

To summarize, starting from an initial state of coherent matter waves at zero momentum, an arrangement of applied interfering laser beams defines a Bravais lattice,
\textit{i.e.}, a set of states 
$\textbf{k}$ connected by allowed 
$\Delta\textbf{k}$ transitions.
The quadratic nature of the free-particle dispersion, which leads to Doppler shifts of the Bragg transitions,
provides a means to control individual Bragg transitions (having a common $\Delta\textbf{k}$ but unique changes in energy) in a spectrally resolved way.
At first order in terms of Bragg transitions, all terms of the relevant tight-binding Hamiltonian are uniquely controllable.
Importantly, the spectroscopic identification of each term of $H_\textrm{eff}$ remains robust as one extends to larger system sizes, as the first-order Bragg resonances are uniformly spaced, such that no spectral crowding ensues.

\subsection{Previous extensions beyond 1D}

Many transport phenomena depend on the dimensionality of the system under consideration, and some effects -- such as the relevance of static complex hopping phases to dynamics -- are entirely absent in one dimension. Thus, there have been previous proposals and attempts to extend MSLs beyond one dimension.

In 1D MSLs, for a given momentum kick leading to a change of atomic wave vector $\Delta\textbf{k}$, the spectrum of relevant Bragg resonances is unique and well-spaced, separated by $8 E_R / \hbar$.
One path to engineering effectively two-dimensional (2D) MSLs was undertaken in \cite{faan2017direct}, by using two sets of Bragg laser beams, having incommensurate wavelengths but oriented along the same spatial axis, to drive unique Bragg transitions that lead to incommensurate changes to the atomic wave vector of $\Delta\textbf{k}_A$ and $\Delta\textbf{k}_B$. Atoms can ``hop'' between modes in an effectively two-dimensional state space. In principle, all of the transitions between sets of neighboring states in this 2D space have a unique resonance frequency. This spectral uniqueness is not robust in practice, however. As atoms move to fill out the effective 2D set of states, they densely fill out the physically 1D span of momentum states \cite{gad2013}. The relevant Bragg spectrum thus becomes denser and denser as the extent of the effective state space is scaled up, and this approach is not suitable for realizing large-scale 2D MSLs.

An analogous approach to building 2D MSLs with unique spectral control of all Bragg transitions was suggested in Ref.~\cite{gad2015}, based on two sets of Bragg lasers aligned along unique but non-orthogonal directions (ideally with a relative angle $\theta$ such that $\cos(\theta)$ is an irrational fraction). This approach leads to a 2D MSL construction, which can in principle allow for unique spectral control of all transitions, however it is also plagued by spectral crowding when scaled to larger system sizes.

\begin{figure*}
\includegraphics[width=1.9\columnwidth]{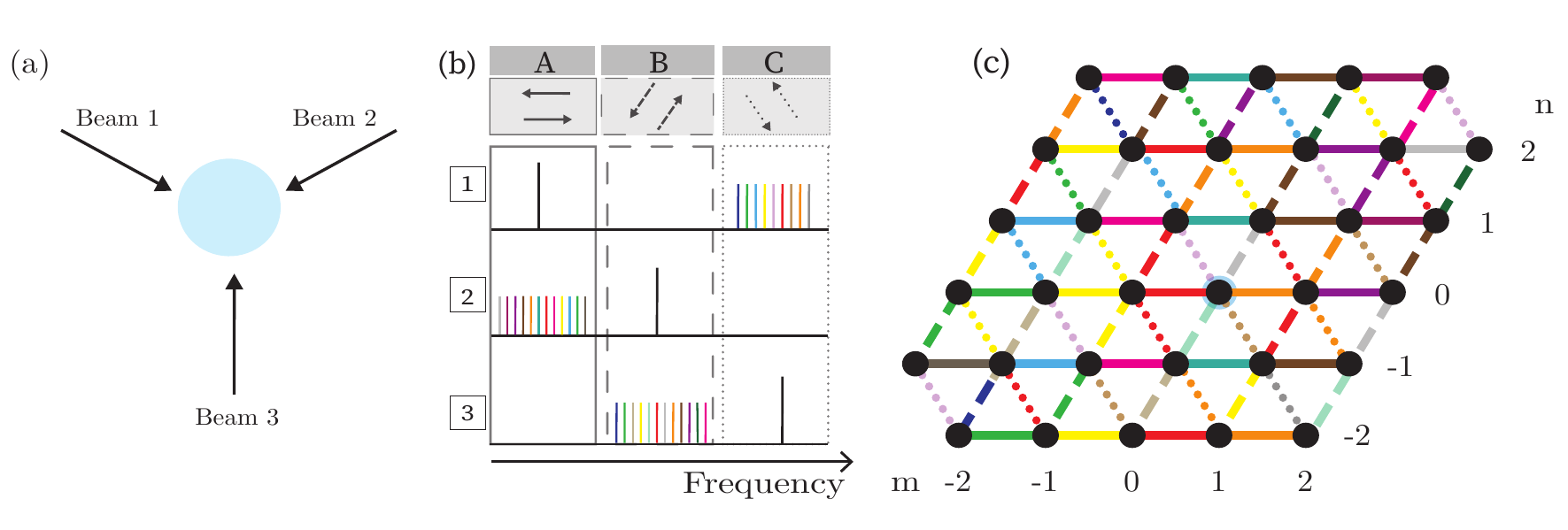}
    \centering
    \caption{\textbf{Two-dimensional momentum state lattices.}
    \textbf{(a)}~The layout of laser beams in real space, aligned in a plane at $120^{\circ}$ relative to one another.
    \textbf{(b)}~The frequency spectra of the beams 1, 2, and 3 as laid out in panel (a). Cross-interference between spectral tones in beams 1 and 2 leads to Bragg transitions along the $A$ direction denoted by the solid horizontal arrows, interference between beams 2 and 3 leads to transitions along the $B$ direction denoted by the dashed arrows, and interference between beams 1 and 3 leads to transitions along the $C$ direction denoted by the dotted arrows.
    \textbf{(c)}~The resulting set of momentum orders (black dots) connected via Bragg transitions from an initial zero momentum state (blue circle) are partially depicted, filling out an effective triangular lattice.
    }
    \label{fig:scheme}
\end{figure*}

Finally, the incorporation of multiple internal states has been suggested or demonstrated for the exploration of non-Abelian tight-binding models~\cite{gad2015}, introducing effective non-Hermitian loss~\cite{Lapp_2019,liang2022}, and as a means of expanding MSLs in an additional (internal) dimension \cite{li2022}. In the context of enabling 2D MSLs, the use of multiple internal states has some unique and positive aspects, however, it is ultimately limited in terms of the system sizes to which it can scale (so far limited to ladder-like systems of two internal states).

\section{MSLs in two dimensions}
\label{2Dmsls}

The simplest approach to forming a two-dimensional momentum state lattice is to simply use two sets of spatially orthogonal and non-interfering Bragg lasers, creating independent MSL-style control for transitions along two orthogonal directions, \textit{e.g.}, $\Delta k_x$ and $\Delta k_y$. Indeed, such an approach would be in some sense very successful at creating scalable 2D MSLs. However, ignoring interactions, the dynamics along the $k_x$ and $k_y$ directions would be entirely \textit{separable}, and this approach would be incapable of exploring any truly two-dimensional phenomena associated with, \textit{e.g.}, gauge fields, non-separable disorder, and genuinely 2D band structures.
This follows from the fact that a given $\Delta k_x$ transition (from some initial value of $k_x$ to $k_x + \Delta k_x$) would have a Bragg resonance frequency that is entirely independent of the $k_y$ coordinate, and likewise for $\Delta k_y$ transitions.
We can contrast such a highly scalable but separable approach to those discussed in Sec.~\pref{1Dmsls}{B}, which preserved the complete and independent control of all relevant Bragg transitions but were severely limited in their scalability. 

Here, we suggest that the optimal approach to building scalable and non-separable 2D MSLs (at least based on first-order Bragg transitions) can be found by considering a compromise between these two extremes. By giving up \textit{complete} control over all Bragg transitions and allowing for resonance conditions (same $\Delta \textbf{k}$ and same $\Delta E$) to repeat at regular intervals, we describe how interesting 2D tight-binding models can be engineered in a fully scalable fashion using momentum state lattices.

In the following, we restrict our discussion to one specific implementation of scalable 2D MSLs based on interfering lasers arranged in a triangular geometry. However, in general, many different laser beam configurations may be used to engineer different kinds of 2D MSLs, with different features as well as restrictions. The general considerations we discuss and the formalism we utilize translate for different 2D MSL arrangements, as well as for extensions to 3D MSLs.

\subsection{Setup, states, and resonance conditions}

Here, we consider the same system of two-level atoms of mass $M$, having a single internal ground (excited) state $\ket{g}$ ($\ket{e}$) with energy $\hbar\omega_{g(e)}$. These two-level atoms and their interaction with a driving electric (laser) field $\textbf{E}$, neglecting spontaneous emission, are described in the dipole approximation by the single-particle Hamiltonian given by Eq.~\ref{dip}.

However, we now consider the electric field $\textbf{E}$ as being based on a laser beam configuration as shown in Fig.~\pref{fig:scheme}{a}. In this layout, the electric field is made up from three co-planar laser fields intersecting at relative angles of 120 degrees~\cite{weit}. In this 2D case, each of the laser fields - labeled as beams $1$, $2$, and $3$ - consists of one isolated frequency component as well as a separate comb of nearby-spaced frequency tones, as depicted in Fig.~\pref{fig:scheme}{b}. Colloquially, we'll refer to the isolated components as the ``carriers'' and the comb-like neighboring tones as ``sidebands.'' As, in the 1D MSL case, the spacing between the components of the ``sideband'' portion will be dictated by the spacing of relevant Bragg resonances, and will again be on the scale of a few~$E_R / \hbar$.
As can be seen from Fig.~\pref{fig:scheme}{b}, the approximate frequencies of the carriers and sideband tones for the three beams are coordinated in such a way as to isolate three pairs of cross-beam interferences that give rise to momentum-changing two-photon transitions, which we will describe in more detail shortly. To note, we assume that the carriers (and sets of sidebands) are mutually separated by a spacing $\delta \gg E_R / \hbar$ (in practice, at the $\delta/2\pi \sim$~few~MHz scale), such that the relevant cross-beam interferences are restricted to those involving intentionally near-coinciding tones (with other terms being neglected by a rotating wave approximation).

For the laser beams labeled by $\sigma \in \{1,2,3\}$, each laser field can be written as 
\begin{align}
    &\mathbf{E}_{\sigma}(\textbf{r},t) = \mathbf{E}^\textrm{c}_{\sigma} \cos(\textbf{k}^\textrm{c}_{\sigma} \cdot \textbf{r} − \omega^\textrm{c}_{\sigma} t + \phi^\textrm{c}_{\sigma}) \nonumber \\
    &\quad + \sum\limits_{j=1}^{N_\sigma} \mathbf{E}^{\textrm{sb},j}_{\sigma} \cos(\textbf{k}^{\textrm{sb},j}_{\sigma} \cdot \textbf{r} − \omega^{\textrm{sb},j}_{\sigma} t + \phi^{\textrm{sb},j}_{\sigma}) \ ,
\end{align}
where the first term relates to the carrier frequency component while the sum relates to the comb of sideband tones.

As in the 1D case, we assume for simplicity that these three laser fields may be approximated as having a spatially homogeneous amplitude over the region of interest and that they have a common polarization along the $z$ axis, \textit{i.e.}, $\mathbf{E}^\textrm{c}_{\sigma} = \textrm{E}^\textrm{c}_{\sigma}\hat{z}$ and $\mathbf{E}^{\textrm{sb},j}_{\sigma} = \textrm{E}^{\textrm{sb},j}_{\sigma}\hat{z}  \ \forall \ j$.
Assuming the electric fields to be nearly monochromatic, we have
\begin{align}
    \textbf{k}^\textrm{c}_{1} &\simeq \textbf{k}^{\textrm{sb},j}_{1} \simeq \textbf{k}_1 \equiv
    k\left(\frac{\sqrt{3}}{2}\mathbf{e}_x - \frac{1}{2}\mathbf{e}_y\right) \\
    \textbf{k}^\textrm{c}_{2} &\simeq \textbf{k}^{\textrm{sb},j}_{2} \simeq \textbf{k}_2 \equiv
    k\left(-\frac{\sqrt{3}}{2}\mathbf{e}_x - \frac{1}{2}\mathbf{e}_y\right) \\
    \textbf{k}^\textrm{c}_{3} &\simeq \textbf{k}^{\textrm{sb},j}_{3} \simeq \textbf{k}_3 \equiv
    k\mathbf{e}_y  \ ,
\end{align}
where $k = 2\pi/ \lambda$ is the wave vector of the laser light having wavelength $\lambda$.

Just like in the 1D case, the near-common single-photon detuning of all the laser fields from atomic resonance ($\Delta \simeq \omega_{eg} - \omega^\textrm{c}_\sigma \simeq \omega_{eg} - \omega^{\textrm{sb},j}_\sigma \ \forall \ \sigma, j$) is assumed
to be much larger than all other relevant terms, including Doppler shifts of magnitude $|p|k/M$ and the resonant Rabi coupling frequencies. This large
single-photon detuning from resonance makes direct population of the atomic excited state $\ket{e}$ negligible. Thus, we may trace out the excited state and effectively describe the system in terms of two-photon Bragg processes that impart momentum to the atoms in the $x-y$ plane.

As described above, and motivated by the structure of the laser spectra shown in Fig.~\pref{fig:scheme}{b}, the interaction of the atoms with the laser fields primarily results in three independent sets of momentum-changing Bragg transitions, which respectively stem from the pairwise interferences of the three beams.
We label these three unique processes by $\Xi \in \{A,B,C\}$, each relating to a unique momentum change $\Delta \textbf{p}_\Xi = \hbar \Delta \textbf{k}_\Xi$. These three discrete Bragg-induced changes to the wave vector of the atoms are defined as
\begin{align}
    \Delta \textbf{k}_A &= \textbf{k}_1 - \textbf{k}_2 = \sqrt{3} k \mathbf{e}_x
    \label{dk1}\\
    \Delta \textbf{k}_B &= \textbf{k}_2 - \textbf{k}_3 = -k\left(\frac{\sqrt{3}}{2}\mathbf{e}_x + \frac{3}{2}\mathbf{e}_y\right)\\
    \Delta \textbf{k}_C &= \textbf{k}_3 - \textbf{k}_1 = k\left(-\frac{\sqrt{3}}{2}\mathbf{e}_x + \frac{3}{2}\mathbf{e}_y\right) \ .
\label{dk3}
\end{align}
The directions of these three classes of allowed Bragg processes ($\pm \Delta \textbf{k}_\Xi$) are also depicted in Fig.~\pref{fig:scheme}{b}.

Assuming that we start with the atomic population at zero momentum, this construction defines a set of possible momentum states labeled as $\ket{m,n}$ that may be populated, having momenta $k_{m,n}=(\sqrt{3} m+\frac{\sqrt{3}}{2}n)\mathbf{e}_x+\frac{3}{2}n \mathbf{e}_y$. The kinetic energies of these $\ket{m,n}$ states are given as
\begin{equation}
    E_{m,n}=3E_R(m^2+n^2+mn)
\end{equation}
Figure~\pref{fig:scheme}{c} shows the set of states that may be populated, filling out a triangular grid in the $k_x - k_y$ plane.
To help guide intuition about the types of lattice structures that may be realizable, we color the transitions for the different $\Delta \textbf{k}_\Xi$ directions according to the corresponding colors of the tones from their generating sideband spectra (the line styles solid, dashed, and dotted respectively relate to the $A$, $B$, and $C$ directions).

The full Hamiltonian of this system is given by
\begin{equation}
    H(t)=\sum_{m,n} E_{m,n}\ket{m,n}\bra{m,n}+\sum_{\Xi}H_\textrm{drive}^{\Xi}(t) \ ,
\label{fullsim2D}
\end{equation}
where $E_{m,n}$ is the kinetic energy of the states forming the momentum state lattice and 
\begin{equation}
    H_\textrm{drive}^{\Xi}(t)= \sum_{m,n} \chi^{\Xi}(t)\ket{m+\Delta m_\Xi,n + \Delta n_\Xi}\bra{m,n} + \text{h.c.} \ .
\end{equation}
Following from Eqs.~\ref{dk1}-\ref{dk3}, the $\Xi$-specific changes to the lattice site indices that result from these Bragg transitions are $\{\Delta m_A,\Delta n_A\} = \{1,0\}$, $\{\Delta m_B,\Delta n_B\} = \{0,-1\}$, and $\{\Delta m_C,\Delta n_C\} = \{-1,1\}$.

\begin{figure*}
    \centering
    \includegraphics[width=1.9\columnwidth]{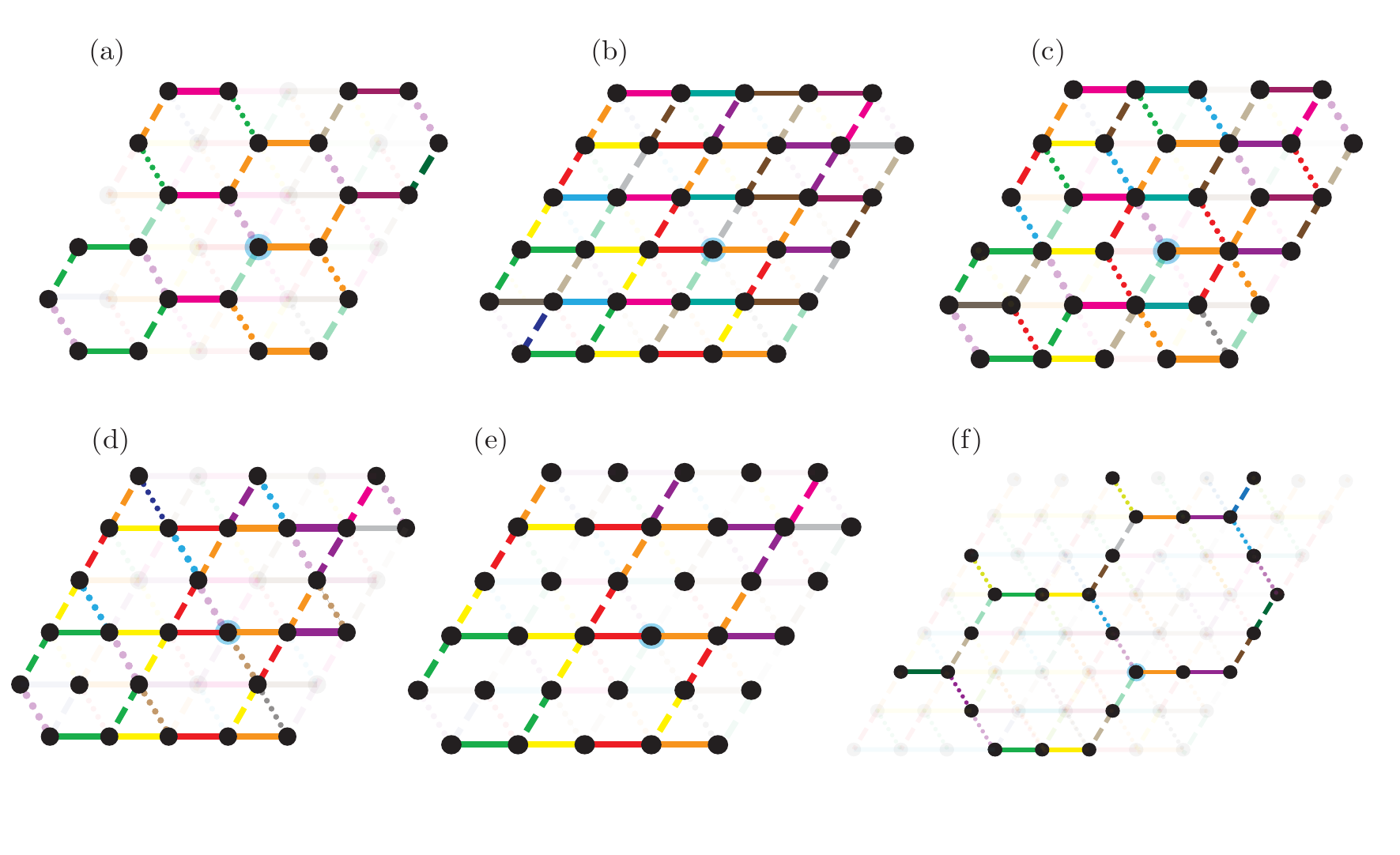}
    \caption{\textbf{Spectral engineering of non-triangular lattices.}
    \textbf{(a-f)}~Illustration of several example lattices that can be engineered by leaving off some resonant tones of the full triangular lattice. The appropriate removal of Bragg driving terms can results in (a)~honeycomb, (b)~square or Hofstadter, (c)~dice, (d)~kagom\'e, (e)~Lieb, and (f)~decorated honeycomb lattices.  Lattices (c) through (f) host flat bands. In the square lattice case of panel (b), specific arrangements of the hopping phases can result in a uniform flux piercing every lattice plaquette, transforming the square lattice into the Hofstadter lattice.}
\label{fig:menag}
\end{figure*}

Here, the driving terms $\chi^\Xi(t)$ are the nearest-neighbor off-diagonal
elements, which assume different compositions for the three $\Delta \textbf{k}$ directions in this scheme. For example, the $\chi^{A}$ spectrum arises from the interference of sidebands of field 2 and the carrier of field 1 and is thus largely defined by the sideband structure of laser beam 2. Explicitly, we have
\begin{equation}
    \chi^{\Xi}(t)= \sum_j \hbar \tilde{\Omega}^\Xi_j e ^{i \phi^\Xi_j} e^{-i \Delta\omega^\Xi_j t} \ .
\end{equation}
In terms of their spectral composition, the $\chi^\Xi$ each host a comb of drive frequencies $\Delta \omega^\Xi_j$, with $\Delta \omega^A_j = \omega^\textrm{c}_1-\omega^{\textrm{sb},j}_2$, $\Delta \omega^B_j =\omega^\textrm{c}_2-\omega^{\textrm{sb},j}_3$, and $\Delta \omega^C_j = \omega^\textrm{c}_3-\omega^{\textrm{sb},j}_1$.
The $\tilde{\Omega}^\Xi_j$ and $\phi^\Xi_j$ terms relate to the individual strengths and phases of these tones, similarly given in terms of the individual single-photon Rabi rates and relative phases of the laser field tones. For example, $\tilde{\Omega}^A_j=\Omega_1^\textrm{c}  \Omega^{\textrm{sb},j}_2/ 2 \Delta$ and $\phi^A_j=\phi^\textrm{c}_1 - \phi^{\textrm{sb},j}_2$.

In the 1D MSL case, each state-to-state Bragg transition had a completely unique resonance condition. This situation is entirely different in 2D. For a given $\Delta \textbf{k}$, there is a family of equivalent transitions - having both the same $\Delta \textbf{k}$ and the same $\Delta E$ - that cannot be separately addressed at first-order. This results directly from the separability of the $(p_x^2 + p_y^2) / 2M$ kinetic energy landscape. For example, for the $A$ transitions, where $\Delta \textbf{k}_A = \sqrt{3} k \mathbf{e}_x$, any transition originating from the same $k_x$ value has the same resonance frequency, independent of the starting $k_y$.

Still, given the set of accessible $\ket{m,n}$ states described above, a consideration of the Bragg resonance conditions along the $A$, $B$, and $C$ directions suggests a large amount of spectroscopic control over 2D MSLs. As in 1D, we may relate these two-photon Bragg resonance conditions to the final-initial state energy difference for the different $\Delta \textbf{k}_\Xi$ transitions, yielding the resonance conditions
\begin{equation}
\hbar \tilde{\omega}_{m,n}^\Xi=E_{m+\Delta m_\Xi,n+\Delta n_\Xi}-E_{m,n}
\equiv 3E_R(\xi+1) \ .
\label{2Dres}
\end{equation}

More explicitly, Eq.~\ref{2Dres} defines the two-photon Bragg resonance conditions for the transitions $\ket{m,n} \leftrightarrow \ket{m+\Delta m_\Xi,n+\Delta n_\Xi}$.
For the different $\Xi$ directions, $A$, $B$, and $C$, the $\xi$ term takes values of $\alpha = 2m+n$, $\beta = -2n-m$, and $\gamma = m-n$, respectively. Thus, the resonance conditions along the $\Xi$ direction are not unique for all sets of $\{m,n\}$, but there are unique sets of resonances $\tilde{\omega}^\Xi_{\xi}$ that identically address all the
$\ket{m,n} \leftrightarrow \ket{m+\Delta m_\Xi,n+\Delta n_\Xi}$ transitions for a given $\xi$ value. For example, the transitions along $A$ from
$\{m,n\} = \ldots, \{1,-2\}, \{0,0\}, \{-1,2\}, \{-2,4\}, \ldots$ are all driven equally by addressing the
$\tilde{\omega}^A_{\alpha = 0}$ resonance.
To note, this particular 2D MSL beam configuration gives rise to a separation of $3E_R/\hbar$ between neighboring first-order Bragg resonances, which is slightly reduced from the $8E_R/\hbar$ separation found in the case of 1D MSLs made from counter-propagating beams.

As before, here we assume that for each of the $\chi^\Xi$ drives each frequency tone $j$ is associated with a unique resonance condition $\xi_j$ up to a small two-photon detuning $\zeta^\Xi_j = \tilde{\omega}^\Xi_{\xi_j} - \Delta \omega^\Xi_j$.
Spectral engineering of an effective time-independent 2D MSL Hamiltonian, $H_\textrm{eff}$, follows by composing frequency tones of the $\chi^\Xi$ drives such that they build up the 2D MSL.
Following a transformation to the interaction picture and a rotating wave approximation, as well as a re-absorption of the $\zeta_j$ terms onto the diagonal by a redefinition of the creation and annihilation operators, the full Hamiltonian in Eq.~\ref{fullsim2D} can be simply recast as a first-order effective 2D MSL Hamiltonian
\begin{equation}
\begin{split}               
H_\textrm{eff}/\hbar &=\sum_{m,n}\varepsilon_{m,n}\hat{c}_{m,n}^{\dagger} \hat{c}_{m,n} \ + \\
&\sum_{\Xi,m,n}J^{\Xi}_{m,n}(e^{i\varphi^\Xi_{m,n}}\hat{c}_{m+\Delta m_\Xi,n+\Delta n_\Xi}^{\dagger}\hat{c}_{m,n}+\text{h.c.})  \ .
    \end{split}
\end{equation}
Just as in 1D, the various terms of this Hamiltonian can be directly related to those of the Bragg drives $\chi^\Xi$. For the example of the $\Xi = A$ transitions, $\tilde{\Omega}^A_j = J^A_{m_j,n_j}$ and $\phi^A_j = \varphi^A_{m_j,n_j}$, where the site indices relate to all transitions satisfying $2m_j + n_j = \alpha_j$ (and similarly for the $B$ and $C$ directions). For the potential terms, to achieve a uniquely defined energy landscape and to avoid residual time-dependencies of the effective Hamiltonian~\cite{Mueller-Escher,Liang-Harp-Hof}, it is separately required that 
$\zeta^\Xi_j = \varepsilon_{m_j+\Delta m_\Xi,n_j+\Delta n_\Xi}- \varepsilon_{m_j,n_j}$ for all the $\Xi$ directions. This condition is specific to multiply-connected lattice graphs.

For visualization, Fig.~\pref{fig:scheme}{c} depicts the base triangular lattice of $\ket{m,n}$ states that are accessible from an initial condensate. Further, the figure indicates transitions belonging to independent $\Xi$ directions and possessing unique resonance frequencies, as identified respectively by the line styles and line colors of the transitions connecting the sites on the graph.

\begin{figure*}[t]
\includegraphics[width=1.9\columnwidth]{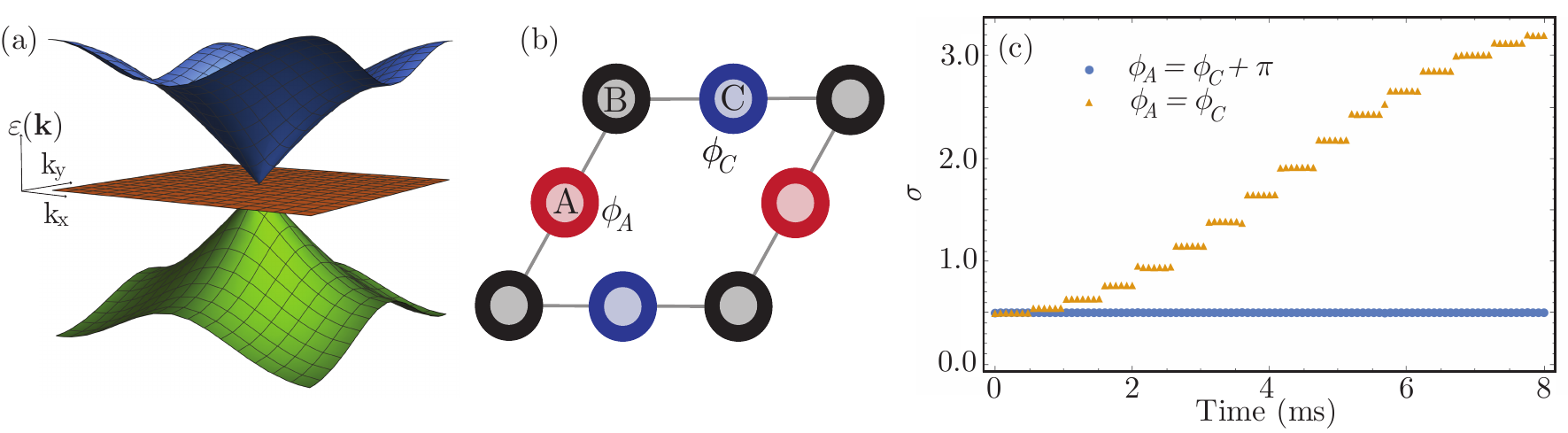}
    \centering
    \caption{\textbf{Frozen dynamics in the Lieb lattice.} \textbf{(a)}~The energy spectrum of the Lieb lattice as a function of the momentum 
    $\textbf{k} = (k_x, k_y)$. \textbf{(b)}~Starting from the same diamond density pattern (equal population amplitudes at the depicted $A$ and $C$ sublattice sites of a given plaquette), selective projection onto either the flat band or dispersive bands can be achieved by controlling the relative phase structure of the wave function at the various site positions.
    For an equal phase at the populated sites ($\phi_A = \phi_C$), dispersing bands are populated, whereas an alternating phase structure ($\phi_A=\phi_C+\pi$) projects solely on to the flat band.
    \textbf{(c)}~Numerical simulations of the dynamics for a hopping rate of $J/2\pi = 202.7$~Hz, starting from the four-site equal phase (orange) and alternating phase (blue) configurations. The dynamics of the standard deviation (in units of the unit-valued lattice site spacing) reveal the difference between frozen flat-band dynamics and population spreading for dispersing band projection.    
    }
    \label{fig:flatband}
\end{figure*}

\subsection{Example first-order Bragg lattices}

Having established a protocol for engineering 2D MSLs and having laid out the general conditions for controlling the unique Bragg resonances, we now provide a few specific examples of different tight-binding models that may be carved out from the triangular template of states.
Starting with this base triangular MSL when all transitions are activated, several alternative tight-binding models may be realized by simply turning off some of the resonant tones, as we show in Fig.~\ref{fig:menag}. To construct the honeycomb lattice, for example, one should only address every third first-order Bragg resonance along each of the $\Xi$ directions. This honeycomb arrangement is displayed in Fig.~\pref{fig:menag}{a}. To note, the choice of whether the various addressed $\xi$ resonances relate to $\{\ldots , 0,3,6,\ldots\}$ or $\{\ldots , 1,4,7,\ldots\}$, \textit{e.g.}, simply results in a discrete shift of honeycomb lattice sites relative to the physical set of $k$ states. This fact is of practical importance, \textit{e.g.}, determining whether or not the zero-momentum condensate does or does not reside on the lattice. 

Figure~\pref{fig:menag}{b} displays an effective square lattice, achieved by addressing all available first-order Bragg tones along two of the $\Xi$ directions labeled by $A$ and $B$ in Fig.~\pref{fig:scheme}{b}.
One reason to highlight this simple lattice variant is that it provides a convenient example of the ability to engineer tunable gauge fields. The unique control over the various tunneling terms of this square lattice, and in particular all of the tunneling terms around the elementary plaquettes, makes possible the construction of fully tunable and uniform Abelian gauge fields. Explicitly, to engineer uniform gauge fields, one should coordinate the tunneling phases as $\phi^A_j=-j\phi$, where $\phi$ is the desired flux value through each plaquette. The flux values can furthermore be randomized to study random flux models~\cite{LeeFisher,ALTLAND1999445} in the square lattice. 

Figure~\pref{fig:menag}{c-f} show a variety of lattices that host dispersionless bands. In these lattices, kinetic frustration is induced by  destructive interference of various tunneling amplitudes, and inter-particle interactions can easily become the dominant energy scale of the system. There have been several past realizations of flat-band optical lattices for ultracold atoms~\cite{taie2015coherent,jo2012}, and their realization in MSLs should present similar as well as new opportunities for exploring the influence of interactions in flat-band systems~\cite{Ozawa-flat,Leung-flat}.

Figure~\pref{fig:menag}{c} displays the dice lattice (or rhombille tiling), which has not been realized by real-space optical lattice techniques to date. As can be seen from the MSL links, the dice lattice is constructed by leaving off every third available Bragg link along each of the $\Xi$ directions.
Figure~\pref{fig:menag}{d} displays the kagom\'e lattice arrangement~\cite{jo2012}, which is the dual of the dice lattice. The kagom\'e lattice can be engineered by simply skipping every other available first-order Bragg resonance along each of the $\Xi$ directions. The Lieb lattice~\cite{taie2015coherent}, as can be seen by inspection of 
Fig.~\pref{fig:menag}{d} and Fig.~\pref{fig:menag}{e}, is produced by using the same $\chi^A$ and $\chi^B$ spectra as used for the kagom\'e lattice, but then simply applying no $\chi^C$ tones.
For all of these canonical flat-band models, the control over state-preparation in MSLs allows for a direct way to initialize atoms in flat bands and explore the influence of atomic interactions.
We also note that the control of 2D MSLs allows for the continuous interpolation between the kagom\'e and Lieb lattices, as well as the potential for adding spin-orbit coupling terms to open up a gap between the flat and dispersing bands~\cite{LiebKagome}.

In addition to the well-known three-band models that possess kinetic frustration (Fig.~\pref{fig:menag}{c-e}), a larger variety of decorated flat-band models~\cite{Dias2015,Dec-sqrt} are also possible to construct by restricting certain MSL sites to have a reduced coordination number, similar to the case of the Lieb lattice. One such decorated lattice, the five-band decorated honeycomb lattice, is depicted in Fig.~\pref{fig:menag}{f}.

Finally, we note that the aforementioned flux control over lattices such as that shown in Fig.~\pref{fig:menag}{b} extends to several of the flat-band models. In particular, a full control over a homogeneous (or inhomogeneous) flux is possible for the Lieb lattice. In contrast, the dice and kagom\'e lattices (as well as the triangular lattice) are restricted in the possible uniform flux arrangements they admit. For the kagom\'e lattice, a flux arrangement (in staggered fashion) is possible only through the triangular plaquettes in the lattice, while for the dice lattice, flux values of $0$, $2 \pi/3$, and $4\pi/3$ are possible. At first order, gauge field engineering is \textit{not at all} possible for the honeycomb and decorated honeycomb lattice due to the parallel pairs of transitions. 
%

\begin{figure*}[t]
\includegraphics[width=1.9\columnwidth]{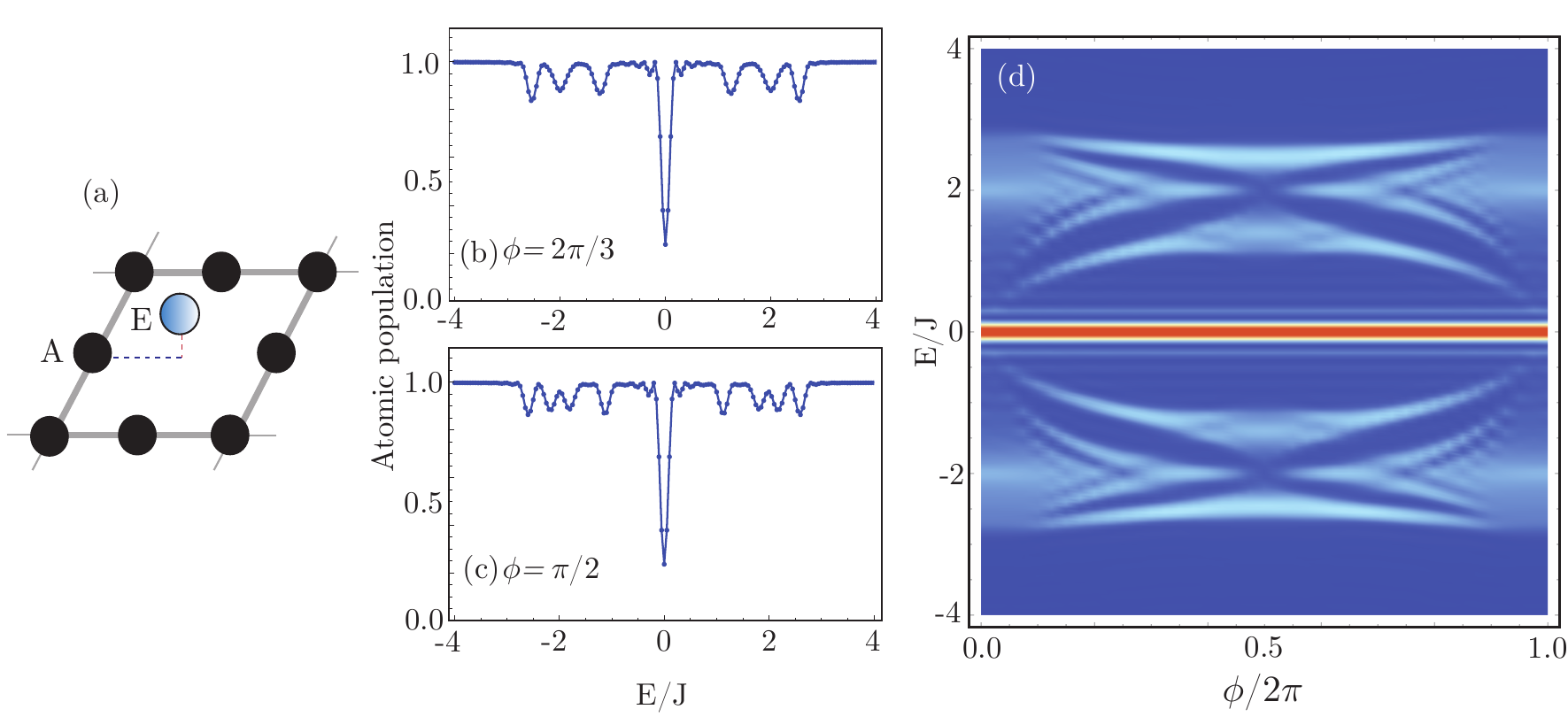}
    \centering
    \caption{\textbf{Loss spectroscopy of flat and topological energy bands.}
    \textbf{(a)}~Tunneling microscopy of the Lieb lattice, monitoring population loss from a weakly coupled probe site (blue, a spectator site) into the
synthetic lattice.
\textbf{(b), (c)}~Probe loss spectra as a function of probe bias $E$, resulting from the resonant weak tunneling of probe atoms into the available lattice eigenstates, shown for lattice flux values of $2\pi/3$ (top) and $\pi/2$ (bottom) per lattice plaquette. The dip features at $E = 0$ relate to the enhanced density of states in the flat band. Features at higher and lower energy than the $E=0$ dip relate to the topological Hofstadter bands~\cite{Goldman-Lieb}. $2q$ loss features can be observed for flux values of $2 \pi /q$, relating to the emergent Hofstadter sub-bands.
\textbf{(d)}~The celebrated Hofstadter butterfly spectrum was obtained from performing numerical simulations such as in (b) for various flux values through the lattice. 
}
    \label{fig:butterfly}
\end{figure*}

\subsection{Frozen dynamics in a flat-band lattice}\label{liebflat}

One of the key new capabilities that would be enabled by 2D MSLs is the achievement of flat-band tight-binding models, \textit{i.e.}, multiply-connected lattices that play host to kinetic frustration.
To note, flat-band phenomenology of the diamond lattice has recently been explored in ladder systems based on 1D MSLs with multiple internal states \cite{li2022}. Extensions to 2D flat-band models based on 2D MSLs would further open up several unique explorations into the interplay of kinetic frustration with atomic interactions, tunable disorder, and non-Hermiticity.

As a concrete demonstration of the ability of 2D MSLs to host this physics, in Fig.~\ref{fig:flatband} we plot simulations of the dynamics of initially localized atomic wave packets in a flat-band Lieb lattice. The Lieb lattice possesses three energy bands (shown in Fig.~\pref{fig:flatband}{a}), two dispersing and one flat, originating from the unit cell structure hosting three sublattice sites (labeled $A$, $B$, and $C$ in Fig.~\pref{fig:flatband}{b}).

The large eigenstate degeneracy of the zero-energy flat band makes it an ideal setting to explore interaction-driven physics~\cite{julku2016} and has motivated studies of mean-field interactions on the dynamics of atoms in real-space Lieb lattices~\cite{taie2015coherent}.
In MSLs, the phase-sensitive control over the properties of an initialized atomic state provides the ability to tunably initialize atoms either within the set of flat-band states or within the dispersing bands.
Fig.~\pref{fig:flatband}{b} and Fig.~\pref{fig:flatband}{c} indicate how this control may be wielded by phase-controlled preparation of an initial state, and how the dispersing dynamics of an equal phase initialized state is sharply contrasted with the frozen dynamics of a staggered phase flat-band state.

Specifically, in Fig.~\pref{fig:flatband}{c} we plot the numerically calculated standard deviation $\sigma$ of the site position operator for dispersing and flat-band states. For an evolving 2D MSL state $\ket{\psi(t)} = \Sigma_{m,n} c_{m,n}(t)\ket{m,n}$, we calculate $\sigma = \sqrt{\sigma_m^2 + \sigma_n^2}$, where $\sigma_{m}$ is calculated as the standard deviation along the $m$ direction, $\sqrt{\bra{\psi(t)} m^2 \ket{\psi(t)}-\bra{\psi(t)} m \ket{\psi(t)}^2}$ (and likewise for $n$). The calculations are based on dynamics under the full Hamiltonian, Eq.~\ref{fullsim2D}, which includes step-like dynamics due to the Floquet nature of the driven MSL system. These calculations assume Rabi rates of $J/2\pi = 202.7$~Hz for the applied Bragg tones, and the stark contrast of the dispersive and flat-band states becomes evident on the few-ms timescale.

\begin{figure*}[t!]
\includegraphics[width=1.9\columnwidth]{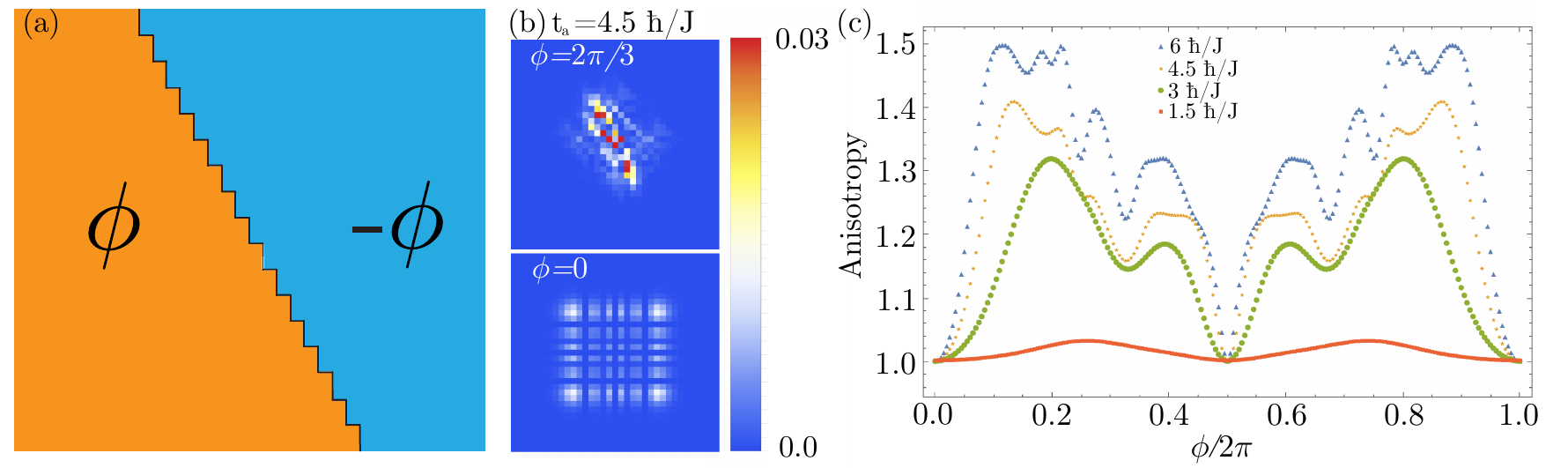}
    \centering   
    \caption{\textbf{Topological boundary states in Hofstadter lattice.}
    \textbf{(a)}~A cartoon showing the ``knight-move" flux boundary implemented in Hofstadter lattice.
    \textbf{(b)}~Density plots of the population distribution after $t_a=4.5$ in units of $\hbar /J$ following initialization at the flux boundary represented in panel (a). The top image relates to the case of $\phi = 2\pi /3$, whereas the bottom plot relates to $\phi = 0$ (and hence no sharp flux boundary).
    \textbf{(c)}~The anisotropy, $\sigma_n/\sigma_m$, of the density distribution following dynamics is plotted as a function of flux through the lattice for various evolution times of 1.5, 3, 4.5, and 6 in units $\hbar/J$. The spread is uniform for $\phi=0,\pi$ while being anisotropic for other flux values.}
    \label{fig:hoff}
\end{figure*}

\subsection{Spectroscopy of the Lieb-Hofstadter lattice}

In the previous section, we looked at the real-space dynamics of the Lieb lattice. Note that in Fig.~\pref{fig:flatband}{c} the dynamics were frozen because the initial state projected onto a flat band and \textit{not} because the system was initialized in a specific eigenstate of the system. Since we started with atoms in a configuration that was not an eigenstate of the system, the subsequent dynamics were by nature out of equilibrium. To motivate the study of eigenstate properties, here we discuss an energy-resolved spectroscopic technique applicable to synthetic dimensions experiments~\cite{SaiPaper}.

We will discuss this spectroscopic technique on the Lieb lattice, as it possesses ``spectator sites'' that are disconnected from the lattice. The presence of such spectator sites makes the Lieb lattice a natural case study for a generalized injection spectroscopy~\cite{Gaebler2010,cheuk,Zhang1,kanungo2022realizing,liang2022,SaiPaper}. Since the eigenspectrum of the Lieb lattice is modified by the presence of flux through the lattice, we study the eigenstates as a function of the flux through the lattice. 

We start with the atomic population at a spectator site, as denoted by the blue lattice site in Fig.~\pref{fig:butterfly}{a}. The energy of this spectator site relative to zero energy is given by $E$. We connect this site to the site on the left, labeled $A$ in Fig.~\pref{fig:butterfly}{a} with a weak tunneling strength of $0.1 J$, where $J$ is the tunneling strength for the lattice. We then observe the population loss from the spectator site as a function of different spectator site energies with respect to the lattice. 

In this numerical experiment, the probe atoms don't just tunnel in and populate site $A$, which is directly linked to the spectator site, but rather tunnel in to the resonant eigenstates that overlap with the $A$ site. The loss rate is proportional to the overlap $|\bra{\psi_{A}}\ket{\psi_i}|^2$, where $\ket{\psi_i}$ are the eigenstates of the lattice. With the help of this scheme, we can probe the energy bands of the Lieb-Hofstadter lattice. More generally, from Fermi's golden rule, the energy bias $E$ of the probe site sets the resonance condition to flow into different lattice eigenstates, with enhanced loss found at such a resonance.

For a particle hopping on a two-dimensional square lattice in the presence of a magnetic field, the spectrum is the celebrated Hofstadter butterfly~\cite{hofstadter1976energy}. In the Hofstadter butterfly, for a magnetic flux of $(p/q) \Phi_0$, where $\Phi_0$ is the flux quantum, the spectrum splits into $q$ bands. The spectrum of a Lieb-Hofstadter lattice consists of two mirrored Hofstadter butterflies, separated by a flat band at $E=0$~\cite{Goldman-Lieb}. Therefore, if we perform a spectroscopic study on the Lieb lattice, we expect to see $2 q$ dips in the spectator site population for a magnetic flux of $(p/q) \Phi_0$, in addition to a dominant dip corresponding to the flat band at $E=0$.

In Fig.~\pref{fig:butterfly}{b} and Fig.~\pref{fig:butterfly}{c}, we plot the population loss of the spectator site in the Lieb lattice geometry for $\phi=2 \pi/3$ and $\phi=\pi/2$ respectively, over 4.5 tunneling times. For both $2\pi/3$ and $\pi/2$ flux values, we observe a sharp dip at $E=0$, indicating the presence of the flat band. The $E=0$ feature persists at all values of $\Phi$. As expected from theory, we see 3 and 4 dips for $2 \pi/3$ and $\pi/2$ flux respectively below and above $E=0$.  Fig.~\pref{fig:butterfly}{d} shows the double Hofstadter spectrum for Lieb lattice, obtained from stitching together multiple line-cuts such as the ones shown in Fig.~\pref{fig:butterfly}{b} and Fig.~\pref{fig:butterfly}{c}.

While this scheme measures the bulk spectra of the Lieb lattice, in the next section we describe a method to probe the edge states of 2D MSLs with a specific example of a square/Hofstadter lattice. 

\subsection{Topological boundary states}

One key feature of MSLs in 1D is the ease of engineering hard-wall open boundary conditions (OBCs), simply by truncating the applied drive spectrum at some chosen Bragg resonance. Such OBCs allow, for example, for the direct investigation of topological boundary states \cite{Stuhl2015,Mancini2015,meier2016,chalopin2020probing,roell2022chiral}.
In 2D MSLs, at least for our described three-beam driving scheme, a similar approach to engineering open boundary conditions leads to a rather meandering edge to the resulting tight-binding model lattices. Thus, a direct interface with ``vacuum'' is not straightforward to implement at first-order, challenging the observation of topological edge modes in 2D. However, by an appropriate arrangement of the hopping phases, 2D MSLs do allow for the engineering of flux interfaces. We can have  regions in the lattice where the flux-per-plaquette of the lattice sharply transitions between two different values.

Here, we explore signatures of topological boundary states at the interface between regions of a Hofstadter lattice hosting two different flux values. Namely, we investigate interfaces between lattices having flux values of $\pm\phi$, thus possessing equivalent band energies and gaps but opposing topological band indexes. In simulations, we probe the topological states appearing at such a flux boundary through nonequilibrium dynamics following initialization near such a boundary in Fig.~\pref{fig:hoff}{a}.

Figure~\pref{fig:hoff}{b} shows a snapshot of the dynamics after $t_a=4.5 \hbar /J $ in the case of a flux boundary (top, $\phi = 2\pi/3$) and a uniform flux (bottom, $\phi = 0$). This time is chosen such that the atomic population is negligible at the edges of a 35 $\times$ 35 lattice, but long enough to see the contrast between the cases of uniform flux and a flux boundary. As one can see in Fig.~\pref{fig:hoff}{b}, in the presence of uniform flux through the lattice, the population spread is isotropic. However, in the presence of a flux boundary, that is, for $\phi=2\pi /3$ in Fig.~\pref{fig:hoff}{a}, the dynamics are anisotropic, and the population spreads favorably along the flux boundary (Fig.~\pref{fig:hoff}{b}). This anisotropy can be explained by the presence of edge states that are gapped from the bulk states for a $\pm 2 \pi/3$ flux boundary, while no such edge states are present for $\phi=0$. Since we initialized the state with no explicit energy selection, the atoms populate both clockwise and counter-clockwise propagating edge modes. 

To further investigate the edge states in the Hofstadter lattice, we plot the anisotropy as a function of flux at different times $t_a$. Here the anisotropy is defined as the ratio of the standard deviation along  $m$ and $n$ directions and is given by $\sigma_n/\sigma_m$, where $\sigma_m$ is calculated as described in sub-section \ref{liebflat}. We see that for $\phi=0$ and $\pi$, the anisotropy is identically equal to 1, with the distribution expanding isotropically for all $t_a$. In contrast, for other flux values that relate to a non-trivial flux boundary, the anisotropy grows as the time $t_a$ increases, and has several finer features appearing at particular values of the flux (\textit{e.g.}, local anisotropy minima at $2\pi/3$, $\pi/2$, etc.).

\section{Further extensions}
\label{extensions}

We've thus far only discussed the simplest possible realizations of 2D MSLs, constructed solely from first-order, two-photon Bragg transitions driven in a uniform manner. Based on the body of work in 1D MSLs, it is natural to expect that some of the most interesting aspects of 2D MSLs will be enabled by capabilities that go beyond this simple description.
Additionally, we've ignored a discussion of how atomic interactions enrich the dynamics in MSL systems.
We now briefly discuss relevant extensions to the 2D MSL construction based on, \textit{e.g.}, spectroscopic control, higher-order Bragg transitions, and atomic interactions.


$ \ $

\noindent\textit{Parameter variation}. 
In the preceding Section, we discussed how different lattices may be realized by either addressing or not addressing Bragg transitions according to specific patterns (cf. Fig.~\ref{fig:menag}). The spectroscopic control of MSLs allows for further variation of the properties - amplitude, phase, detuning - of \textit{addressed} transitions. Indeed, controlled variations of the phases of Bragg tones is what allows for the engineering of artificial gauge fields, such as in the Hofstadter model (and, \textit{e.g.}, the Lieb-Hofstadter model). We also discussed how patterns of gauge fields, particularly flux boundaries, can be engineered. Beyond this, controllably random gauge fields~\cite{LeeFisher,ALTLAND1999445} can be introduced by applying random variations to the tunneling phases along multiple Bragg axes.

Spatial parameter variation can also be introduced in the amplitudes and detunings of the Bragg tones, which enables control over hopping amplitudes and site-energy potentials of the resulting MSLs. Such control would be important, for example, in the study of localization physics in pseudodisordered quasicrystals.

Finally, in addition to spatial variations of the Hamiltonian parameter values, time-dependent changes can also be readily achieved. Already in 1D MSLs, this control has been important for explorations of Floquet systems~\cite{Xiao2020} and the realization of time reflection~\cite{dong2023quantum}, and many interesting scientific directions will be made accessible by the extension of this control to 2D MSLs.

$ \ $

\noindent\textit{Higher-order processes}.
In one-dimensional MSLs, higher-order Bragg processes that act as beyond nearest-neighbor hopping terms have played an important role in expanding the kinds of lattices that can be engineered. Specifically, four-photon (second-order Bragg) transitions enable the introduction of multiply-connected pathways in one-dimensional MSLs, thus making possible the design of artificial gauge fields~\cite{faan2018,gou2020}. In 2D MSLs, where the introduction of gauge fields is realized naturally at first order (with respect to the Bragg processes), higher-order Bragg terms can still play a useful role. One simple example is by promoting the honeycomb structure of Fig.~\pref{fig:menag}{a} to a topological Haldane lattice~\cite{Haldane88,Jotzu2014}.

To achieve this extension from the honeycomb lattice to a Haldane model requires that different (parallel) second-order Bragg transitions that share the same net $\Delta\textbf{k}$ and $\Delta E$ should be implemented with different complex hopping phases. While this is forbidden for first-order Bragg transitions, it can be accomplished at second-order by using distinct combinations of coordinated first-order processes for each of the desired second-order links. Indeed, if the individual elements of a given second-order Bragg process relate to unique directions of momentum change, such as a $\Delta k_A$ and $\Delta k_B$, then their tandem four-photon transition (with a rate $\sim \Omega_a \Omega_b / 2\Delta_i$, with $\Delta_i$ their characteristic detuning from a common intermediate state and $\Omega_{a,b} \ll \Delta_i$ their respective first-order Rabi rates) can be uniquely associated with both the initial $k_x$ and $k_y$ values of the atoms. In contrast to the case of first-order Bragg transitions, this allows for a unique control of \textit{all} second-order Bragg transitions in a way that is fully momentum-selective. To note, while this in principle allows for the unique design of essentially \textit{any} tight-binding model (having sites that reside on a given grid of $\textbf{k}$ states), the implementation at second-order requires operation at greatly reduced tunneling rates, and is thus less robust than the first-order implementation.

Finally, one particular second-order process worth highlighting is the introduction of local effective loss terms~\cite{Lapp_2019,gou2020,Chen2021,liang2022}, achieved by coupling momentum states to either an auxiliary reservoir of momentum states or a distinct internal level that can be selectively removed by resonant light~\cite{Lapp_2019}. The local control of such loss processes has opened up MSL systems to the study of non-Hermitian behaviors, such as the non-Hermitian skin effect~\cite{liang2022}, and the extension to 2D MSLs promises to facilitate the exploration of even richer non-Hermitian phenomena.

$ \ $

\noindent\textit{Interactions}. Over the past decade, the exploration interaction effects in 1D MSLs have provided a means to explore beyond-single-particle physics~\cite{An-Corr,Xie-Walk,Guan-NonExp,chen2021grosspitaevskiiequation,An-nonlinear,Wang-Interactions} and provide a route towards direct interaction-based squeezing of atomic momentum modes~\cite{An-Corr}.
Similar diagonal (configuration-preserving) interaction terms should enrich the physics of 2D MSLs, in particular allowing for connections to the physics of mean-field solitons as recently explored in photonic systems~\cite{Muk-nonlinear,Jurgensen2021,Jurg-theory,Mostaan2022,Jurgensen2023}

Two-dimensions MSLs also offer a qualitatively new aspect of interactions as compared to 1D MSLs -- the relevance of state-changing four-wave mixing processes that conserve both energy and momentum, which in the language of MSLs relate to correlated tunneling (anti-tunneling) processes of atom pairs~\cite{Rolston2002}.

$\ $

\noindent\textit{3D MSLs}. Finally, we briefly remark that our overall discussion of the generalization of 1D MSL techniques to 2D MSLs (expounding on just one particular implementation) can be naturally extended to the generation of 3D MSLs, based on a set of two-photon Bragg transitions that fill out a discrete set of momentum states, characterized by nearest-neighbor momentum-change vectors that have projections onto three orthogonal axes. Just as the extension to 2D opens up the exploration of artificial gauge fields and kinetic frustration, the subsequent extension to 3D MSLs would open up new opportunities to explore, \textit{e.g.}, tailored Weyl materials, dimensionality-dependent localization phenomena, and more. Short of a full extension to a implementation in three physical momentum directions, the extension to layer-like geometries based on multiple internal states could be achieved, similar to recent extensions of 1D MSLs~\cite{li2022}.

$ \ $

\section{Conclusion}
\label{conc}

In conclusion, we have proposed a straightforward path to extending the control of synthetic momentum state lattices to two dimensions and beyond. Such techniques naturally give rise to models with controllable gauge fields, kinetic frustration, and additional control capabilities relevant to quantum simulation and the exploration of topological and localization phenomena. As has been the case in 1D MSLs, further extensions based on added internal states, higher-order Bragg transitions, and other added features promise to open up the realization of many new exciting transport phenomena beyond what has been explicitly discussed herein.

$ \ $

\section{Acknowledgements}
This material is based upon work supported by the Air Force Office of Scientific Research under Grant No.~FA9550-21-1-0246 and the AFOSR MURI program under agreement number FA9550-22-1-0339.
We would like to thank Kuan-Sen Lin, Jedediah Pixley, Xiye Hu, Tao Chen, Fangzhao Alex An, Eric Meier, and Jackson Ang'ong'a for discussions.

\bibliographystyle{apsrev4-1}
\bibliography{msl2d}

\end{document}